\documentstyle[12pt]{article}

\newcommand{\resection}[1]{\setcounter{equation}{0}\section{#1}}

\newcommand{\EQ}{\begin{equation}}
\newcommand{\EN}{\end{equation}}
\newcommand{\A}{\begin{array}}
\newcommand{\E}{\end{array}}
\newcommand{\EA}{\begin{eqnarray}}
\newcommand{\EE}{\end{eqnarray}}

\newcommand{\Z}{{\sf Z\!\!Z}}
\newcommand{\goto}{\rightarrow}
\newcommand{\hs}{\hspace{1mm}}

\newcommand{\ep}{\varepsilon}
\newcommand{\D}{{\rm d}}
\newcommand{\de}{\delta}
\newcommand{\th}{\theta}
\newcommand{\om}{\omega}
\newcommand{\siml}{\raisebox{-.6ex}{$\stackrel{<}{\displaystyle{\sim}}$}}
\newcommand{\simg}{\raisebox{-.6ex}{$\stackrel{>}{\displaystyle{\sim}}$}}
\newcommand{\ind}{\scriptstyle}

\newcommand{\NP}{Nucl. Phys. }

\begin{document}
\setcounter{page}{0}
\topmargin 0pt
\oddsidemargin 5mm
\renewcommand{\thefootnote}{\fnsymbol{footnote}}
\newpage
\setcounter{page}{0}
\begin{titlepage}
\begin{flushright}
November 1991
\end{flushright}
\vspace{0.8cm}
\begin{center}
{\large {\bf MULTIPLE CROSSOVER PHENOMENA AND SCALE HOPPING IN
          TWO DIMENSIONS }}

\vspace{1.6cm}
{\large
Michael L\"{a}ssig} \footnote{
Electronic mail: iff299@DJUKFA11, lassig@iff011.dnet.kfa-juelich.de}

\vspace{0.9cm}
{\em Institut f\"ur Festk\"orperforschung \\
     Forschungszentum J\"ulich \\
     5170 J\"ulich, Germany } \\
\end{center}
\vspace{1.3cm}

\setcounter{footnote}{0}

\begin{abstract}

We study the renormalization group for nearly marginal perturbations of a
minimal conformal field theory $M_p$ with $p \gg 1$. To leading order in
perturbation theory, we find a unique one-parameter family of ``hopping
trajectories'' that is characterized by a staircase-like renormalization group
flow of the $C$-function and the anomalous dimensions and that is related to a
recently solved factorizable scattering theory \cite{AlZam.roam}. We argue that
this system is described by interactions of the form
$ t \phi_{(1,3)} - \bar t \phi_{(3,1)} $. As a function of the {\em relevant}
parameter $t$, it undergoes a phase transition with new critical exponents
simultaneously governed by all fixed points $M_p$, $M_{p-1}$, ..., $M_3$.
Integrable lattice models represent different phases of the same integrable
system that are distinguished by the sign of the {\em irrelevant} parameter
$\bar t$.

\end{abstract}

\end{titlepage}

\newpage
\renewcommand{\thefootnote}{\arabic{footnote}}

\resection{Introduction}

The simplest scale-invariant field theories in two dimensions are the series of
minimal models $M_p$ ($p=3,4,\dots$) \cite{BPZ}, which describe the universal
$(p-1)$-critical behavior of Landau-Ginzburg theories with a single bosonic
field and polynomial interactions \cite{Zam.LG}. It is a difficult and widely
open  problem to reveal the embedding renormalization group (RG) scenario of
these fixed points, which determines the universal behavior off criticality as
well as crossover phenomena.  An important aspect of this problem is that in
two dimensions the theory has an infinite number of integrals of motion not
only  at the RG fixed points, where it is conformally invariant \cite{BPZ}, but
on a larger submanifold of theory space. The precise extent of
this manifold of integrability is unknown, but it does contain some
perturbations of a critical point $M_p$ by a single scaling field
\cite{Zam.int}.

If such a perturbation is {\em relevant}, it will either induce a crossover  to
another critical point of lower criticality or lead to purely  massive infrared
behavior. In the latter case, the exact factorizable $S$-matrix can be
conjectured \cite{Zam.int}, which in principle determines all scaling functions
associated to that RG trajectory. At least  some properties of this scaling
regime have indeed been predicted, such as universal finite-size effects
\cite{Luescher,AlZam.Potts,AlZam.yl,Klassen.TD,AlZam.RSOS,YZ.YL_Ising,TIM,SCATT}
and amplitude relations \cite{AMPL}.  The only known example of an integrable
crossover from the model $M_p$ to another critical point is generated by the
perturbation \EQ
{\cal L} = {\cal L}_p^{\star} + t_p \phi_{p (1,3)} \hs,
\label{13}
\EN
where $\phi_{p (1,3)}$ is the weakest relevant scaling field, i.e. the field
with the smallest positive RG eigenvalue, and the (dimensionful) coupling
constant $t_p$ is positive. This crossover changes the order of criticality by
one to $M_{p-1}$; this has been shown perturbatively for large values of $p$
\cite{Zam.cth,LudCardy.pert}, by supersymmetry arguments for the crossover from
the tricritical to the critical Ising model (the case $p=4$) \cite{KMS}, and
more recently by thermodynamic Bethe ansatz methods for general values of $p$
\cite{AlZam.TIM}. Hence the manifold ${\cal C}_p$ of $(p-1)$-criticality is
nested into all manifolds of lower criticality, as one would exspect from a
mean-field analysis of the Landau-Ginzburg picture:   ${\cal C}_p \subset {\cal
C}_{p-1} \subset \dots \subset {\cal C}_3$.  In contrast to mean-field
arguments, however, crossovers changing the order of criticality by more than
one are induced by fine-tuned linear combinations of all scaling fields that
are even under spin reversal, and it is not clear if any of the interpolating
field theories are integrable.

The manifold of integrability also contains the leading {\em irrelevant}
scaling perturbation of the critical theories $M_p$,
\footnote{The bar does not denote complex conjugation.}
\EQ
{\cal L} = {\cal L}_p^{\star} -\bar t_p \phi_{p (3,1)}
\label{31}
\EN
at least to first order in perturbation theory \cite{Zam.int,AlZam.RSOS}. While
generically a linear combination of two integrable perturbations does not
generate an integrable field theory off criticality
\footnote{
The two-dimensional Ising model was studied recently \cite{IS} in the entire
scaling region ${\cal L} = {\cal L}^{\star} + h \phi_{(1,2)} + t \phi_{(1,3)}$,
which is spanned by two integrable perturbations. The finite-size spectrum of
the transfer matrix obtained by the conformal truncation method
\cite{YZ.YL_Ising,TIM,SCATT} did not show any sign of integrability except for
$h=0$ or $t=0$. },
the perturbations
(\ref{13}) and (\ref{31}) share infinitely many integrals of motion so that
even an arbitrary linear combination
\EQ
{\cal L} = {\cal L}_p^{\star} + t_p \phi_{p (1,3)} -\bar t_p \phi_{p (3,1)}
\label{mix}
\EN

should still be integrable \cite{AlZam.RSOS}. As will be argued below, this
fact is connected to the existence of nontrivial integrable lattice models in
two dimensions. The presence of nonzero irrelevant coupling constants in
lattice models can drastically alter their crossover behavior: since the
$(p-1)$-critical lattice model is  characterized by a point on ${\cal C}_{p}$
different from the fixed point $M_p$, the variation of a thermodynamic
parameter causing the continuum theory $M_p$ to  cross over to  ${\cal C}_{p'}$
$(p' < p)$ need not be tangent to  ${\cal C}_{p'}$ at that point, which leaves
the perturbed lattice model on a less critical manifold  ${\cal C}_{p''}$
($p''< p'$) or in a massive phase.

This paper studies such crossover phenomena by analyzing the renormalization
group flow in the neighborhood of a minimal model $M_p$ for $p \gg 1$, where a
subset of the scaling fields (including $\phi_{p (1,3)}$ and
$\phi_{p (3,1)}$) become nearly  marginal. A perturbation of $M_p$ can then
be described by the Lagrangian

\EQ
         {\cal L} = {\cal L}_p^{\star} + \sum_i U_{p}^i \,\Phi_{p i} \hs;
\label{lagr}
\EN
the running coupling constants $U_{p}^i$ and their conjugate fields  $\Phi_{p
i}$ are defined by an expansion in the parameter $\ep = 4 /(p+1)$ (which is the
RG eigenvalue of $\phi_{p (1,3)}$) \cite{Zam.cth,LudCardy.pert}. The
$\ep$-expansion can be trusted in a neighborhood $U_{p}^i = O(\ep)$ of $M_p$,
which contains infinitely many other fixed points $M_{p'}$. This is an
important difference to the usual $\ep$-expansion about the upper critical
dimension, where only two fixed points are at a distance of $O(\ep)$. But at
least on the trajectory linking $M_p$ and $M_{p-1}$, this $\ep$-expansion has
been shown to be a consistent RG scheme in minimal subtraction to  $O(\ep^2)$
\cite{CTH}.

To leading order in perturbation theory, we find in particular a unique
one-parameter family of {\em hopping trajectories} $\check U_{p}^i(\th,\th_0)$
(where $\th$ is the RG ``time'' varying along each trajectory and $\th_0$
labels
the trajectories). They come close to each fixed
point  $M_p$ and are {\em self-similar} in the following sense:
\EQ
\check U_{p-1}^i (\th+\th_0, \th_0) = \check U_{p}^i (\th, \th_0) \hs.
\label{sim}
\EN
We identify these trajectories with the one-parameter family of integrable
trajectories that Al.B. Zamolodchikov recently found by solving the
thermodynamic Bethe ansatz for a simple factorizable scattering theory
containing a single type of massive particles \cite{AlZam.roam}. The flow of
the $C$-function $C(\th,\th_0)$ along these trajectories is computed and seen
to follow the characteristic staircase pattern that interpolates between the
central charges $c_p$. A similar pattern is found for the flow of the anomalous
dimensions $x^{(i)}(\th,\th_0)$. We shall argue that the one-parameter family
$M(\th_0)$ of integrable field theories defined by these $S$-matrices
is described by a Lagrangian of the form (\ref{mix}) where both coupling
constants $t_p$ and $\bar t_p$ are positive.

The same family of field theories can be considered for negative  values of
$\bar t_p$, where the RG trajectories behave very differently:  they come close
to only two neighboring fixed points $M_p$ and $M_{p-1}$ and should describe an
integrable system in a $(p-2)$-phase coexistence region. This is very likely to
be the eight-vertex  solid-on-solid model of Andrews, Baxter and Forrester
(ABF) \cite{ABF} in the  scaling region of the so-called regime IV. The RG
analysis thus establishes an intimate connection between this model and
Zamolodchikov's system.

As a function of the relevant temperature-like parameter  $t_{p}$, the system
undergoes a second order phase transition with a rather intricate critical
behavior.  For  $t_{p} < 0$ and any value of $\bar t_p$,  it is governed by the
{\em single} fixed point  $M_p$. For $t_{p} > 0$ and $\bar t_p < 0$, {\em two}
neighboring fixed points determine the exponents; the RG confirms the scaling
ansatz proposed by Huse \cite{Huse} to explain the exponents in regime  IV of
the ABF-model.  For $t_{p} > 0$ and $\bar t_p > 0$, they are determined by {\em
all} fixed points  $M_{p}, M_{p-1}, M_{p-2},\dots, M_3$ visited by the hopping
trajectories.

This paper is organized as follows. In sect. 2, we write down the RG equations
and determine some useful symmetry properties. Sect. 3 discusses the hopping
trajectories. Sect. 4 describes the various phase coexistence regions and the
critical behavior as a function of $t_{p}$. Sect. 5 contains a discussion of
the results.

\resection{First-order renormalization about a minimal model $M_p$}

To leading order in perturbation theory, the RG equations about the fixed point
$M_p$ can be written in the form \cite{Zam.cth,Cardy.LH}
\footnote{
{}From now on, the index $p$ will be suppressed where no ambiguities can
arise.}
\EQ
\frac{\D}{\D \th} U^i =
y_{\hs j}^{i} U^j - \pi C_{\hs jk }^{i} U^j U^k \hs,
\label{RG}
\EN
where $x^{(i)} = (2 - y^{(i)})$ are the anomalous dimensions and
$C_{\hs jk }^{i}$ the structure constants
of the scaling operators
$\phi_i = \Phi_i(U\!=\!0)$, and
$y_{\hs j }^{i} = y^{(i)} \de^{i}_{\hs j}$.
In Eq. (\ref{RG}), the indices run over all scaling fields that transform as
scalars under rotations, including
pure derivative fields $\partial_z \partial_{\bar z} \phi_i$. Hence the RG acts
on a space of coupling constants whose dimensionality exceeds that
of the thermodynamic space of the system. At any point $U$, there are linear
combinations of the fields $\Phi_i(U)$ that are proportional to the pure
derivative fields $\partial_z \partial_{\bar z} \Phi_i(u)$ and hence not
conjugate to any thermodynamic parameter; these fields generate redundant
\cite{Wegner.DG} directions in coupling constant space.

The RG equations can be restricted to the ``thermal" couplings
that preserve the $\Z_2$-symmetry of $M_p$ under spin reversal. Further
simplifications arise in an expansion in the parameter
$\ep \equiv y^{(1,3)} = 4/(p+1)$. Such an expansion is possible since
both the structure constants
and the scaling dimensions are analytic in $\ep$. For $\ep \goto 0$, the
scaling fields $\phi_i$ in the lower left corner
of the Kac table (shown in fig.~1) are spectrally separated:
\newline (i) the primary fields $\phi_{(m,n)}$ with $|m-n| \leq 1$
have dimension $x^{(m,n)} \leq 1/2 + O(\ep)$,
\newline (ii) the primary fields $\phi_{(n, n \pm 2)}$
and the (conveniently normalized \cite{Zam.cth}) descendant fields
$\tilde{\phi}_{(n,n)}
\equiv (x^{(n,n)})^{-1} \partial_z \partial_{\bar z} \phi_{(n,n)}$
have dimension $2 \mp O(\ep)$ and $2 - O(\ep^2)$, respectively, and
\newline (iii) all other fields have dimension $\geq 5/2 + O(\ep)$
\footnote{The spectral separation breaks down for $n = O(p)$, but for the
solutions of the RG equations to be discussed in the sequel, the couplings of
these fields are exponentially suppressed.}.
\newline Hence the couplings $U^{(n,n \pm 2)}$ and
$\tilde{U}^{(n,n)}$ become marginal in this limit, while all other
couplings remain strictly  relevant or irrelevant. To leading order in $\ep$,
the system of equations (\ref{RG}) can be  truncated consistently to the nearly
marginal couplings $U^i = O(\ep)$, the other couplings remain of
$O(\ep^2)$. A convenient  rescaling
$U^i(\th) = \ep u^i(\ep \tau) / (\pi C^{\ind (1,3)}_{\ind (1,3)(1,3)}) $
then brings the RG equations into the form
\EQ
\frac{\D}{\D \tau} u^i =
\gamma^i_{\hs j} u^j - c^i_{\hs jk} u^j u^k + O(\ep)
\label{RGsc}
\EN
with $\gamma^i_{\hs j} = \lim_{\ep \goto 0} (y^i_{\hs j} / \ep)$ and
$c^i_{\hs jk} =
\lim_{\ep \goto 0} (C^i_{\hs jk} / C^{\ind (1,3)}_{\ind (1,3)(1,3)})$.

These equations determine in particular the {\em renormalizable manifold}
${\cal R}_p$ of $M_p$, i.e. the set of all trajectories
\EQ
u_{p}^i(\tau) \hspace{1cm} \mbox{with } \hspace{3mm}
u_{p}^i(\tau) \goto 0 \mbox{ for } \tau \goto - \infty
\EN
and the {\em critical manifold} ${\cal C}_{p}$ of $M_p$, i.e. the set of
all trajectories
\EQ
u_{p}^i(\tau) \hspace{1cm} \mbox{with } \hspace{3mm}
u_{p}^i(\tau) \goto 0 \mbox{ for } \tau \goto + \infty \hs,
\label{cmft}
\EN
modulo the redundant couplings (see fig. 2). The $(p-p')$-dimensional
{\em crossover manifold}
\EQ
M_{p,p'} = {\cal R}_p \cap {\cal C}_{p'}
\EN
describes the $(p-p'-1)$-parameter family of field theories whose ultraviolet
asymptotics is determined by $M_p$ and whose infrared behavior is determined
by $M_{p'}$. The simplest such solution is the unique trajectory $M_{p,p-1}$
\cite{Zam.cth},
\EQ
u^{(1,3)}(\tau) = \frac{\mbox{exp}(\tau - \tau_m)}{1 + \mbox{exp}(\tau -
 \tau_m)} \hs,
\hspace{1cm} u^i(\tau) = 0 \hspace{0.3cm} \mbox{for } i \neq (1,3) \hs,
\label{u13}
\EN
where $\tau_m$ is a free parameter. This trajectory interpolates between $M_p$
and the infrared fixed point $u^{(1,3)}_{\star} = 1$ associated to $M_{p-1}$.

Under a simultaneous RG time reversal and basis change involving a reflection
about the diagonal of the Kac table,
\EQ
\tau \goto -\tau \hs, \hspace{1cm}
\phi_{(m,n)}=\phi_i \goto - \phi_{\bar i} = -\phi_{(n,m)} \hs,
\EN
the equations (\ref{RGsc}) remain invariant since
$\gamma^{\bar i}_{\hs \bar j} = -\gamma^i_{\hs j}$ and
$c^{\bar i}_{\hs \bar j \bar k} = c^i_{\hs jk}$.
Hence to every RG trajectory $u^i(\tau)$, there is a conjugate trajectory
$\bar u^i(\tau) = -u^{\bar i}(-\tau)$,
and to every fixed point $u^i_{\star}$,
there is a conjugate fixed point
$\bar u^i_{\star} = -u^{\bar i}_{\star}$. For example, the trajectory
conjugate to $M_{p,p-1}$ interpolates between $M_p$ and the ultraviolet fixed
point $u^{(3,1)}_{\star} = -1$ associated to $M_{p+1}$. Of
particular importance in the sequel will be the self-conjugate trajectories,
which satisfy
\EQ
u^i(\tau) = -u^{\bar i}(-\tau + \tau_1)
\EN
for some value of $\tau_1$.

In a small neighborhood of the trajectory $M_{p,p-1}$, the RG
equations can be linearized in the other couplings
$v^i \equiv u^i \ll 1 \hs (i \neq (1,3))$. The equation for $u \equiv
u^{(1,3)}$
then decouples and $u(\tau)$ is given by Eq. (\ref{u13}); the equations for
$v^i$ take the form
\EQ
\frac{\D}{\D \tau} v^i = \gamma^i_{\hs j} (u(\tau)) v^j \hs,
\label{RGlin}
\EN
where $\gamma^i_{\hs j}(u)$ factorizes into (3$\times$3)-matrices
\EQ
\gamma^{(n)} =
\left ( \A{ccc}
   \frac{n+1}{2}         &            0               &          0
\\
        0                &            0               &          0
\\
        0                &            0               &  - \frac{n-1}{2}
\E \right )
- 2 u
\left ( \A{ccc}
   \frac{n+3}{n+1}       & \frac{n-1}{n+1} \left (
                           \frac{n+2}{n} \right)^{1/2}&          0
\\
\frac{n-1}{n+1} \left (
\frac{n+2}{n} \right)^{1/2} & \frac{4}{n^2 - 1}    &\frac{n+1}{n-1} \left (
                                                   \frac{n-2}{n} \right)^{1/2}
\\

        0                & \frac{n+1}{n-1} \left (
                           \frac{n-2}{n} \right)^{1/2}&   \frac{n-3}{n-1}
\E \right )
\label{ylin}
\EN
acting on the triplets of couplings
\EQ
v^{(n)} = \left ( \A{c} u^{(n,n+2)}                \\
                        \tilde{u}^{(n,n)}          \\
                        u^{(n,n-2)}                \E \right )
\EN
with $n=3,5,7,\dots$.

A basis of solutions of Eq. (\ref{RGlin})
is given by the trajectories with the definite RG time reversal symmetry
\EQ
(u_{p}(\tau_m - \tau), v_{p}(\tau_m -\tau)) =
(\bar u_{p-1}(\tau_m + \tau), \pm \bar v_{p-1}(\tau_m + \tau)) \hs.
\label{sym}
\EN
where $u_{p}(\tau)$ is given by Eq. (\ref{u13}). For each value of $n$, there
is precisely one linearly independent even solution $v_{p}^{(n)+}$ and two
linearly independent odd solutions $v_{p}^{(n)-}$ and $v_{p}^{(n)0}$. The
couplings $v_{p}^{(n)0}$ are conjugate to the pure derivative fields
$\partial_z \partial_{\bar z} \Phi_{(n,n)}(u)$ of of $M_{p,p-1}$ and play a
redundant r\^ole.

For the asymptotic behavior of the trajectories as $\tau \goto -\infty$
(i.e. $u_p \equiv u \goto 0$), there are three possibilities. For every value
of $n$, there is one linearly independent solution of (\ref{RGlin}),
\EQ
v_p^{(n) r} (u) = u^{\frac{n+1}{2}} \left [
\left ( \A{c} 1 \\ 0 \\ 0 \E \right ) + u
\left (\A{c} - \frac{2(n+3)}{n+1}                                            \\
             - \frac{4(n-1)}{(n+3)(n+1)} \left (\frac{n+2}{n} \right )^{1/2} \\
             0
\E \right ) + O(u^2) \right ] \hs,
\label{vrel}
\EN
defining a renormalizable trajectory $ (u_p(\tau), v_p^{(n) r}(u(\tau)))$ and
one solution
\EQ
v_p^{(n) 0} (u) =
\left ( \A{c} 0 \\ 1  \\ 0 \E \right ) + u
\left (\A{c}   \frac{4}{n+1} \left (\frac{n+2}{n} \right )^{1/2} \\
             - \frac{8}{n^2 - 1}                                 \\
             - \frac{4}{n-1} \left (\frac{n-2}{n} \right )^{1/2} \\
\E \right ) + O(u^2)
\label{vred}
\EN
defining a redundant trajectory. Any
solution that is linearly independent from (\ref{vrel}) and (\ref{vred})
describes a theory that is nonrenormalizable about $M_p$.
Conversely, there is one linearly independent solution $v_p^{(n) c} (u)$ that
defines a trajectory in ${\cal C}_{p-1}$; any solution that is linearly
independent of $v_p^{(n) c} (u)$ and $v_p^{(n) 0} (u)$ is of lower criticality.

\resection{Self-similar hopping trajectories}

In this section, we study the RG flow of a self-conjugate perturbation
of the fixed point $M_p$,
\EQ
u_{p}^i(\tau \! = \! 0) = \bar u_{p}^i (\tau \! = \! 0) \ll 1 \hs,
\label{sc}
\EN
corresponding to a point in theory space that is much closer to $M_p$ than any
of the other fixed points $M_{p'}$. We define the parameter
\EQ
s \equiv u_p^{(1,3)} \bar u_p^{(1,3)}  > 0 \hs.
\label{s}
\EN

It is easy to verify that there is a unique one-parameter family
$\hat u_{p}^i (\tau, s)$ of trajectories that satisfy the conditions
(\ref{sc})
and $\hat u_{p}^i (\tau, s) \subset {\cal C}_{p-1}$, i.e. $\hat v_{p}$ can
be expanded in the basis of triplets $v_{p}^{(n) c}$ and $v_{p}^{(n) 0}$.
\footnote{
For a given value of $s$, one has
$\hat u_p^{(1,3)} = s^{1/2} = - \hat u_p^{(3,1)}$
by Eqns. (\ref{sc}) and (\ref{s}).
The remaining couplings are recursively determined by the equations
\[
\hat v_{p}^{(n)}(0,s) =
a_{(n) c}(s) v_{p}^{(n) c}(u \!=\! s^{1/2})
+ a_{(n) 0}(s) v_{p}^{(n) 0}(u \!=\! s^{1/2})
\]
and the self-conjugacy conditions
\[
\hat u_{p}^{(n,n+2)}(0,s) = - \hat u_{p}^{(n+2,n)}(0,s)
\hspace{0.5cm} \mbox{and} \hspace{0.5cm}
\hat{\tilde u}_{p}^{(n,n)}(0,s) = 0 \hs.
\]
Analogous recursion relations hold for
the trajectories $\check u_{p}^i (\tau,s)$ below.
Notice that for both
families and every value of $n$, the ratio of the relevant and the irrelevant
coupling $u_{p}^{(n,n+2)}(0,s) / u_{p}^{(n,n-2)}(0,s)$ in the $n$th
triplet goes to 0 as $s \goto 0$, hence the couplings with higher $n$ are
strongly suppressed for small $s$ and $\tau$.}
Self-conjugacy then dictates
$\hat u_{p}^i (\tau,s) \subset {\cal R}_{p+1}$
as well and therefore
\EQ
\hat u_{p}^i (\tau,s) \subset M_{p+1,p-1} \hs.
\label{ucross}
\EN
Any trajectory in $M_{p+1,p-1}$ with a given value of $s$ differs from
$\hat u_{p}^i (\tau,s)$ only by spurious couplings $v_{p}^{(n) 0}$, hence
these trajectories (shown in fig. 3a) span the two-dimensional crossover
manifold $M_{p+1,p-1}$.

Consider now the one-parameter family $\check u_{p}^i (\tau,s)$  of
trajectories that satisfy the condition (\ref{sc}) and are  even in the sense
of Eq. (\ref{sym}) up to spurious couplings, i.e.  $v_{p}$ can be expanded in
the basis of triplets $v_{p}^{(n) +}$ and $v_{p}^{(n) 0}$. The trajectory
$\check u_{p}^i (\tau,s)$ is self-similar (see fig. 3b) after a scaled RG time
$\tau_0 = \ep \th_0 \simeq \log(1/s)$ for small $s$:
\EQ
\check u^i_{p-1} (\tau_0,s) =
\bar{\check u}_{p}^i (0,s) = \check u_{p}^i(0,s) \hs.
\label{uself}
\EN
It comes close to each fixed point $M_{p'}$ in the time interval
\EQ
\A{c}
(p-p'-\frac{1}{2}) \th_0 \,\siml\, \th \,\siml\, (p-p'+\frac{1}{2}) \th_0
\E
\label{time}
\EN
and up to a minimum distance given by the parameter $s_{p'} = s(1 + O(\ep))$,
whereafter it hops to the next lower fixed point. It is again easy to check
that up to spurious couplings, this is the only  self-similar trajectory for
that value of $s$. Thus the trajectories
$\check u_{p}^i (\tau, s \!=\! s(\th_0))$ define a unique one-parameter family
of field theories $M(\th_0)$.

The RG flow of the $C$-function \cite{Zam.cth}
\EQ
C(u_{p}^i) = c_p + \frac{3 \ep^3}{16}
 (-3 \gamma_{ij} u_{p}^i u_{p}^j + 2 c_{ijl} u_{p}^i u_{p}^j u_{p}^l)
 + O(\ep^4)
\EN
for the theory $M(\th_0)$ satisfies
\EQ
C(\th, \th_0) - c_p = C(\th + \th_0, \th_0) - c_{p-1} + O(\ep^4)
\label{csim}
\EN
and in particular for integer $k$
\EQ
\A{c}
C(k \th_0, \th_0) = c_{p-k} + O (\ep^4) \\
C( (k + \frac{1}{2}) \th_0, \th_0) =
c_{p-k} - \frac{1}{2} ( c_{p-k} - c_{p-k-1} ) + O(\ep^4)
\E
\label{cfix}
\EN
by Eq. (\ref{uself}). A step of this self-repeating staircase pattern for
several values of $\tau_0$ is shown in fig. 4, which was obtained by numerical
integration of Eq. (\ref{RGlin}).
\footnote{Eqns. (\ref{ucross}) and (\ref{uself}) also indicate the possibility
that there exists a two-parameter family of trajectories
$u_{p}^i (\tau, \hat s, \check s)$ with
$u_{p}^i (\tau, \hat s, \check s) \goto \check u_{p}^i (\tau, \check s)$
as $\hat s \goto 0$ and
$u_{p}^i (\tau, \hat s, \check s) \goto \hat u_{p}^i (\tau, \hat s)$
as $\check s \goto 0$, their $C$-function being a staircase pattern where all
steps have approximately the same length except the step at $c_p$, which is
shorter.}

The anomalous dimensions $x^{(i)}$, i.e. the eigenvalues  of the matrix
\EQ
2 \de^i_{\hs j} - y^i_{\hs j}(u^l_{p}) =
2 \de^i_{\hs j} - y^i_{\hs j (p)} + 2 \ep c^i_{\hs j l} u_{p}^l + O(\ep^2)
\EN
show a very similar pattern. For example, the spectral flow  associated to the
second subdiagonals of the Kac table satisfies
\EQ
\A{c}
x^{(n,n+2)} (\th + \th_0, \th_0) = x^{(n-2,n)} (\th, \th_0) + O(\ep^2) \\
x^{(n,n-2)} (\th + \th_0, \th_0) = x^{(n+2,n)} (\th, \th_0) + O(\ep^2)
\E \EN
and in particular for integer $k$
\EQ
\A{l}
x^{(n,n+2)} (k \th_0, \th_0) = x_{p}^{(n-2k,n+2-2k)} + O(\ep^2)
\hspace{1cm} (2k < n) \\
x^{(n,n-2)} (k \th_0, \th_0) = x_{p}^{(n+2k,n-2+2k)} + O(\ep^2)
\hspace{1cm} (2k > -n)  \hs,
\E \EN
and similar equations hold for the other fields.

Thus the field theories $M(\th_0)$ behave under RG transformations in a
strikingly similar way to the one-parameter family of  integrable systems with
a single type of massive particles characterized by  the factorizable
$S$-matrix \cite{AlZam.roam}
\EQ
S(\rho,\th_0) =
\frac{\sinh \rho - i \cosh 2 \th_0}{\sinh \rho + i \cosh 2 \th_0} \hs,
\label{S}
\EN
written in terms of the Lorentz-invariant rapidity difference $\rho$.  Since
the self-similarity  (\ref{uself}) is unique to the the theories $M(\th_0)$, we
are lead to identify them with this type of integrable system. It is plausible
that the hopping trajectories $\check u_{p}^i (\tau,s)$ describe
integrable systems since the bare Lagrangian
$ (\partial / \partial s) \sum_i (\check U_{p}^i \Phi_{p i}) (0,s)|_{s=0}$
is of the form (\ref{mix}) with $t_{p} > 0$ and $\bar {t}_{p} > 0$,
but it is difficult to make such a statement precise within the $\ep$-expansion
since these trajectories are nonrenormalizable about any minimal model
$M_{p'}$.

The following scaling argument indicates, however, that the Lagrangian
\EQ
{\cal L}_p = {\cal L}_p^{\star} + \sum_i t_p^i \phi_{p i}
\EN
for the theories $M(\th_0)$ is precisely (\ref{mix}) for any value of $p$.
We define the dimensionless scaling variables
\EQ
s_p^i \equiv t_p^i      t_p^{\om_p^i}         \hs,  \hspace{1cm}
  \bar s_p^i \equiv t_p^i \bar t_p^{\bar \om_p^i}     \hs,
\EN
where $\om_p^i$ and $\bar \om_p^i$ are the crossover exponents
\EQ
    \om_p^i \equiv - \frac{ y_p^{(i)} }{ y_p } \hs, \hspace{1cm}
\bar \om_p^i \equiv - \frac{ y_p^{(i)} }{ \bar y_p } \hs,
\EN
with
\EQ
      y_p \equiv y_p^{(1,3)} = \frac{4}{p+1} \hs,  \hspace{1cm}
  \bar y_p \equiv y_p^{(3,1)} = - \frac{4}{p} \hs.
\label{y1331}
\EN
The theory $M_{p,p-1}$ has $s_p^i = 0$ for $i \neq (1,3)$ and
$\bar s_{p-1}^i = 0$ for $i \neq (3,1)$, while for any finite value of
$\th_0$, the theory $M(\th_0)$ must have irrelevant couplings
\EQ
s_p^i \ll 1  \hspace{1cm} \mbox{ for } \th_0  \gg 1 \hs,
\label{sirr}
\EN
since its ultraviolet behavior differs from $M_p$, and relevant couplings
\EQ
s_{p-1}^i \ll 1  \hspace{1cm} \mbox{ for } \th_0  \gg 1 \hs,
\label{srel}
\EN
since its infrared behavior differs from $M_{p-1}$. There is an analytic
mapping between the two sets of couplings (\ref{sirr}) and (\ref{srel}), which
is just a coordinate transformation on theory space \cite{CTH}. The most
relevant coupling $s_{p-1}^{{\rm rel}}$ of the set  (\ref{srel})  and the most
irrelevant coupling $s_p^{{\rm irr}} \sim s_{p-1}^{{\rm rel}}$ of the set
(\ref{sirr})  determine the logarithmic scale intervals (the RG time intervals)
in which the theory $M(\th_0)$ is governed by the fixed points $M_p$ and
$M_{p-1}$, respectively:
\EQ
\mbox{exp} (\Delta \th_p)
\sim ( s_p^{{\rm irr}} )^{  1 / y_p^{{\rm irr}}  }
\gg 1 \hs, \hspace{1cm}
\mbox{exp} (\Delta \th_{p-1})
\sim ( s_{p-1}^{{\rm rel}} )^{ -  1 / y_{p-1}^{{\rm rel}} }
\gg 1 \hs.
\EN
Zamolodchikov's solution \cite{AlZam.roam} of the thermodynamic Bethe ansatz
equations for the $S$-matrix (\ref{S}) says that Eq. (\ref{time}) is valid
beyond perturbation theory, i.e. $\Delta \th_p = \Delta \th_{p-1} = \th_0$.
This
dictates
\EQ
\frac{1}{y_p^{{\rm irr}} }
= - \frac{ 1 }{ y_{p-1}^{{\rm rel}} } \hs,
\EN
which can be satisfied only if $t_p^{{\rm irr}} \sim t_p^{(3,1)}$ and
$t_{p-1}^{{\rm rel}} \sim t_{p-1}^{(1,3)}$, and hence  $y_p^{\rm irr}
= -4/p = -y_{p-1}^{{\rm rel}}$ by Eq. (\ref{y1331}). Repeating the
argument for $p'= p+1$ then fixes the form of the Lagrangian
(\ref{mix}). Hence from the Lagrangian point of view, the scale hopping of the
theories $M(\th_0)$ is caused by an intricate interplay of the relevant field
$\phi_{(1,3)}$ and the irrelevant field $\phi_{(3,1)}$ under the
renormalization group.

The scaling parameter $s_p \equiv \bar t_p t_p^{\om_p}$
(with $\om_p \equiv \om_p^{(3,1)} = (p+1)/p $) can be expressed by
\EQ
s_p = g_p^{\om_p} \bar g_p \,\mbox{exp}( \bar y_p \th_0 )
\label{exp_th}
\EN
in terms of $\th_0$ and the dimensionless coupling constants
\EQ
g_p = t_p \xi_{p,p-1}^{y_p} \hs,              \hspace{1cm}
\bar g_p = \bar t_p \xi_{p+1,p}^{\bar y_p} \hs.
\EN
Here $\xi_{p,p-1}$ and $\xi_{p+1,p}$ denote the crossover length scales of
$M_{p,p-1}$ and $M_{p+1,p}$, respectively. Unlike the running couplings,
$g_p$ and $\bar g_p$ are measurable parameters related to universal amplitude
relations \cite{AMPL}. By comparing the solution of the thermodynamic Bethe
ansatz with conformal perturbation theory, they can be computed to arbitrarily
high accuracy \cite{AlZam.Potts,AlZam.TIM}. To leading order in the
$\ep$-expansion, one obtains
\footnote{
Hence to this order, $s_p$ coincides with the parameter $s$ defined in
Eq. (\ref{s}).}
\EQ
g_p = \bar g_p = \frac{\ep}{\pi C^{(1,3)}_{(1,3)(1,3)}} + O(\ep^2) \hs.
\EN

\resection{Phase coexistence and critical behavior}

The integrability of the theories given by the Lagrangian (\ref{mix})
should not depend on the sign of the two coupling constants $t_{p}$ and
$\bar t_{p}$. However, the behavior of the RG trajectories and hence the
long-distance structure crucially depends on these signs: the four
one-parameter families of field theories
\EQ
\A{cccc}
M_p^{++}(s_p) = M(\th_0(s_p)) \hs,  &
M_p^{+-}(s_p)                 \hs,  &
M_p^{-+}(s_p)                 \hs,  &
M_p^{--}(s_p)                 \hs, \E
\EN
labeled by the scaling parameter $s_p$ and the signs of $t_p$ and $\bar t_p$,
describe the system in different thermodynamic phases which we discuss below.
The qualitative RG scenario and the resulting phase diagram in the $(t_{p},
\bar t_{p})$-plane are shown in fig. 5. As a function of the  relevant
parameter $t_{p}$, the system undergoes a second order phase transition whose
exponents  depend on the phase.

\subsection{The theories $M_p^{++}$}

For $t_{p} >0$ and $\bar t_{p} >0$, the solutions of the RG equations are the
self-similar hopping trajectories $\check u_{p}^i$. Following such a trajectory
down to $M_3$ shows that also $t_3 > 0$; the system is in a  disordered
high-temperature phase. As $t_p \goto 0$, these trajectories come arbitrarily
close to all fixed points $M_{p'}$. This implies that the leading thermodynamic
singularities are governed by the fixed points $M_{p'}$ with $p' \leq p$, while
the other fixed points contribute corrections to scaling. Exact critical
exponents for these theories will be reported in a forthcoming publication
\cite{HOPPING2}.

\subsection{The theories $M_p^{+-}$}

The trajectories for $t_{p} >0$ and $\bar t_{p} < 0$ are obtained by
analytically continuing the  solutions $\check u_{p}^i = (\check u_{p}, \check
v_{p})$ of regime 1 to $(\check u_{p}, -\check v_{p})$ in the neighborhood of
$M_{p,p-1}$, i.e. at times $0 \siml \th \siml \th_0(s)$; these solutions are
still even under RG time reversal according to Eq. (\ref{sym}) up to spurious
couplings. At times $\th \simg \th_0(s)$, they approach the trajectory
$M_{p-1}^{(-)}$ generated by the integrable perturbation (\ref{13}) of
$M_{p-1}$ with $t_{p-1} < 0$ which describes the system in a low-temperature
region of $p-2$ coexisting phases. It is likely that this  one-parameter family
of solutions shares the same infrared behavior. Hence  they are very different
from the self-similar trajectories: they come close to only two fixed points
$M_p$ and $M_{p-1}$ and run away in both time limits,
\EQ
\A{cc}
\bar u_{p}(\th,s) \goto -\infty  \mbox{ as } \th \goto -\infty \hs, &
u_{p-1}(\th,s) \goto -\infty     \mbox{ as } \th \goto +\infty \hs. \E
\EN

An ordered phase above the critical ``temperature'' $t_p = 0$ is not to be
expected for the continuum theory, but it does occur in regime IV of the
ABF lattice model. It is easy to show that the above renormalization group
picture indeed reproduces the correct order parameter exponents known from the
exact solution. The behavior of the trajectories indicates that the theories
$M_p^{+-}(s_p)$ are characterized by two length scales, the crossover scale
$\xi_{p,p-1}$ and the inverse mass $\xi$. Their asymptotic temperature
dependence is given by
\EQ
\xi_{p,p-1} \sim t_p^{ - 1/y_p }
\EN
and
\EQ
\frac{ \xi }{ \xi_{p,p-1} } \sim \mbox{exp} \th_0
\sim t_p^{ \om_p / \bar y_p }
= t_p^{ - \om_p / y_{p-1}  } \hs,
\EN
as follows from Eqns. (\ref{exp_th}) and (\ref{y1331}). For the leading
singular behavior of the order parameters
\EQ
\langle \phi_{p (n,n)} \rangle \sim
\xi_{p,p-1}^{ - x_p^{(n,n) } }
\left ( \frac{ \xi }{ \xi_{p,p-1} } \right )^{ - x_{p-1}^{(n,n)} } \hs,
\EN
we obtain therefore
\EQ
\langle \phi_{p (n,n)} \rangle
\sim t_p^{ \beta_p^{(n,n)} + \om_p \beta_{p-1}^{(n,n)} }
\EN
with $\beta_p^{(n,n)} = x_p^{(n,n)} / y_p $, which is precisely Huse's result
\cite{Huse}. We conclude that the theories $M_p^{+-}$ describe the ABF
model in the scaling region of regime IV. This explains the phase structure as
a consequence of the same interplay of $\phi_{1,3}$ and $\phi_{3,1}$ that
causes the scale hopping of the theories $M_p^{++}$.

\subsection{The theories $M_p^{-+}$}

For $t_{p} <0$ and $\bar t_{p} > 0$, the trajectories are conjugate to those of
$M_p^{+-}$ and in fact just those of $M_{p+1}^{+-}$; they describe the system
in the $(p-1)$-phase coexistence region.  At times $ -\th_0(s) \siml \th \siml
0$, they come close to the fixed point $M_{p+1}$, the trajectory $M_{p+1,p}$,
and the fixed  point $M_p$; at large times, they run away,
\EQ
\A{cc}
\bar u_{p+1} \goto -\infty \mbox{ as } \th \goto -\infty \hs, &
       u_{p} \goto -\infty \mbox{ as } \th \goto +\infty \hs. \E
\EN
The critical behavior as $t_p \goto 0$ is governed by the fixed point $M_p$,
with corrections to scaling due to the irrelevant operator $\phi_{p (3,1)}$.

\subsection{The theories $M_p^{--}$}

For $t_{p} <0$ and $\bar t_{p} < 0$ (hence $s>0$), one exspects solutions  that
are again self-conjugate, $u_{p}^i(\th,s) = -\bar u_{p}^i(-\th,s)$, and
describe the system in the $(p-1)$-phase coexistence region. They come close
only to one fixed point $M_p$, and run away at large times,
\EQ
\A{cc} \bar u_{p}^{(1,3)} \goto -\infty \mbox{ as } \th \goto -\infty \hs, &
            u_{p}^{(1,3)} \goto -\infty \mbox{ as } \th \goto +\infty \hs. \E
\EN
The theories $M_p^{--}$ should describe the ABF models in the scaling region of
regime III. The critical behavior is governed by $M_p$, but the corrections to
scaling are of opposite sign compared to the theories $M_p^{-+}$.

\resection{Discussion}

We have studied perturbations of a minimal conformal field theory $M_p$ by a
linear combination of the scaling fields $\phi_{(1,3)}$ and $\phi_{(3,1)}$.
This generates four one-parameter families of massive integrable field
theories $M_p^{++}(s_p)$, $M_p^{+-}(s_p)$, $M_p^{-+}(s_p)$ and $M_p^{--}(s_p)$,
which are labeled by the signs of the two coupling constants and the
dimensionless scaling parameter $s_p$, and describe the system in different
phases off criticality.

The disordered high-temperature phase corresponds to the theories
$M_p^{++}(s_p)$, which are related to Zamolodchikov's scattering theory
(\ref{S}), and show a novel behavior under the renormalization group: the
trajectories come close to many fixed points $M_{p'}$ for a certain RG time
interval $\th_0(s_p)$, whereafter they hop to the next fixed point $M_{p'-1}$.
The correlation functions of these theories are characterized by a multitude of
crossover length scales $\xi_{p,p-1}$; any two subsequent such scales have
the same ratio $\xi_{p-1,p-2} / \xi_{p,p-1} = {\rm e}^{\th_0}$.

To leading order in an $\ep$-expansion, we have shown that the RG equations
have indeed a unique one-parameter family of solutions with this behavior,
which is tied to the simultaneous presence of relevant and irrelevant coupling
constants with  scaling dimensions of $O(\ep)$.

The theories $M_p^{+-}(s_p)$ and $M_p^{--}(s_p)$ are argued to describe the
scaling region of the ABF lattice models in regime III and IV, respectively.
It would be interesting to study the corrections to scaling in these exactly
solved models. Are all nonanalytic corrections due to irrelevant operators
in the family of $\phi_{(3,1)}$? This would severely restrict the possible
lattice effects. And is it possible to find lattice models with the leading
irrelevant coupling of opposite sign, that would hence be in
Zamolodchikov's phase?

\noindent {\sl Note added:} After the draft of this paper had been completed, I
received a copy of ref. \cite{Klassen.Dflow}, where Zamolodchikov's  $S$-matrix
for antiperiodic boundary conditions is associated to the $D$ series of minimal
models. The issue of boundary conditions deserves further study. The Lagrangian
description suggests that integrable systems with scale hopping trajectories
should exist in the $A$ and $D$ series.

\section*{Acknowledgments}

I enjoyed helpful discussions with John L. Cardy, who also gave the manuscript
a critical reading, and with Reinhard Lipowsky.


\newpage

\section*{Figure Captions}

\begin{enumerate}

\item Positions of the nearly marginal thermal operators in the Kac table of a
unitary  minimal model $M_p$ for $p \gg 1$. The operators
$\phi_{(n,n+2)}$ are  relevant, while the operators
$\phi_{(n,n-2)}$ and $\tilde \phi_{(n,n)}$ are irrelevant.

\item Special solutions of the RG equations in the vicinity of $M_{p,p-1}$
(schematic).
(a) A trajectory in ${\cal R}_p$.
(b) A trajectory in ${\cal C}_{p-1}$.
(c) A redundant trajectory.

\item Self-conjugate trajectories in the vicinity of the fixed point $M_p$
(schematic).
(a) A trajectory in $M_{p+1,p-1}$.
(b) A self-similar trajectory.

\item The $C$-function $C(\tau,\tau_0)$ of the unique self-similar trajectory
$\check u_{p}^i(\tau,s(\tau_0))$: a step in the staircase pattern for   $\tau_0
= 3.2$, $3.6$, $4.0$ and $4.4$ (solid lines). For larger values of  $\tau_0$,
the steps get more pronounced as the solutions tend towards the limit
trajectory $M_{p,p-1}$ (long-dashed line).

\item
(a) The RG flow in the vicinity of $M_p$. A self-similar trajectory of regime 1
(solid line) visits all fixed points, trajectories in regime 2 or 3
(long-dashed lines) visit two fixed points, and a trajectory in regime 4
(short-dashed line) visits only one fixed point.
(b) The resulting phase diagram in the $(t_{p}, \bar t_{p})$-plane.

\end{enumerate}

\end{document}